\documentclass[11pt,letterpaper]{article}
\usepackage{color,ifpdf,latexsym,graphicx,url}
\usepackage{CJKutf8}

\title{\vspace{5ex}\smash{\includegraphics[width=\textwidth]{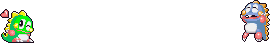}}\vspace{-9ex}%
  Bust-a-Move/Puzzle Bobble is NP-Complete}
\author{%
  Erik D. Demaine%
    \thanks{MIT Computer Science and Artificial Intelligence Laboratory,
      32 Vassar St., Cambridge, MA 02139, USA, \protect\url{edemaine@mit.edu}}
\and
  Stefan Langerman%
    \thanks{Directeur de recherches du F.R.S.--FNRS, D\'epartment
      d'Informatique, Universit\'e Libre de Bruxelles, Brussels, Belgium,
      \protect\url{stefan.langerman@ulb.ac.be}}
}
\date{}

\newif\ifabstract
\abstracttrue
\newif\iffull
\ifabstract \fullfalse \else \fulltrue \fi

\usepackage
  [breaklinks,bookmarks,bookmarksnumbered,bookmarksopen,bookmarksopenlevel=2]
  {hyperref}
{\makeatletter \hypersetup{pdftitle={\@title}}}

\advance \topmargin by -\headheight
\advance \topmargin by -\headsep
\evensidemargin \oddsidemargin
{\makeatletter
 \gdef\xxxmark{%
   \expandafter\ifx\csname @mpargs\endcsname\relax 
     \expandafter\ifx\csname @captype\endcsname\relax 
       \marginpar{xxx}
     \else
       xxx 
     \fi
   \else
     xxx 
   \fi}
 \gdef\xxx{\@ifnextchar[\xxx@lab\xxx@nolab}
 \long\gdef\xxx@lab[#1]#2{\textbf{[\xxxmark #2 ---{\sc #1}]}}
 \long\gdef\xxx@nolab#1{\textbf{[\xxxmark #1]}}
}

{\makeatletter \gdef\fps@figure{!htbp}}

\let\realbfseries=\bfseries
\def\bfseries{\realbfseries\boldmath}

\newtheorem{theorem}{Theorem}

\let\epsilon=\varepsilon

\begin{document}
\maketitle

\begin{abstract}
  We prove that the classic 1994 Taito video game, known as Puzzle Bobble or
  Bust-a-Move, is NP-complete.  Our proof applies to the perfect-information
  version where the bubble sequence is known in advance, and it
  uses just three bubble colors.
\end{abstract}

\begin{quote} \raggedleft
\sl ``A girl runs up with somethin' to prove. \\
\sl So don't just stand there. Bust a move!'' \\
\rm --- Young MC \cite{YoungMC}
\end{quote}

\section{Introduction}

Erik grew up playing the action platform video game \emph{Bubble Bobble}
(\begin{CJK}{UTF8}{min}バブルボブル\end{CJK}),
starring cute little brontosauruses Bub and Bob,%
\footnote{Spoiler: if you finish Bubble Bobble in super mode in co-op,
  then the true ending reveals that Bub and Bob are in fact human boys,
  transformed into brontosauruses by the evil whale Baron Von Blubba
  \cite{RetroGamer}.}
on the Nintendo Entertainment System.
(The game was first released by Taito in 1986, in arcades
\cite{Arcade-Museum-Bubble-Bobble}.)
Some years later (1994), Bub and Bob retook the video-game stage with
the puzzle game \emph{Puzzle Bobble}
(\begin{CJK}{UTF8}{min}パズルボブル\end{CJK}),
known as \emph{Bust-a-Move} in the United States
\cite{Arcade-Museum-Puzzle-Bobble,Wikipedia-Puzzle-Bobble}.
This game essentially got Stefan through his Ph.D.: whenever he needed a
break, he would play as much as he could with one quarter.
To celebrate the game's 21-year anniversary,
we analyze its computational complexity,
retroactively justifying the hours we spent playing.

In Puzzle Bobble, the game state is defined by a hexagonal grid, each cell
possibly filled with a \emph{bubble} of some color.  In each turn,
the player is given a bubble of some color, which can be fired in any
(upward) direction from the pointer at the bottom center of the board.
The fired bubble travels straight, reflecting off the left and right walls,
until it hits another bubble or the top wall, in which case it terminates at
the nearest grid-aligned position.
If the bubble is now in a connected group of at least three bubbles of the
same color, then that group disappears (``pops''), and any bubbles now
disconnected from the top wall also pop.

Here we study the perfect-information (generalized) form of Puzzle Bobble.
We are given an initial board of bubbles and the entire sequence of
colored bubbles that will come.  The goal is to clear the board using the
given sequence of bubbles.  (The actual game has an infinite, randomly
generated sequence of bubbles, like Tetris
\cite{Breukelaar-Demaine-Hohenberger-Hoogeboom-Kosters-Liben-Nowell-2004}.)
The game also has a falling ceiling, where all bubbles descend every fixed
number of shots; and if a bubble hits the floor, the game ends.
We assume that the resolution of the input
is sufficiently fine to hit any discrete cell that could be hit by an
(infinitely precise) continuous shot.  (This assumption seems to hold in
the original game, so it is natural to generalize it.)

\begin{theorem}
  Puzzle Bobble is NP-complete.
\end{theorem}

Membership in NP is easy: specify where to shoot each of the $n$ given bubbles.
The rest of this paper establishes NP-hardness.

Our reduction applies to all versions of Puzzle Bobble.
Viglietta \cite{Viglietta-2014-games-conf} proved that Puzzle Bobble 3 is
NP-complete, by exploiting ``rainbow'' (wildcard) bubbles.
Our proof shows that this feature is unnecessary.

\section{NP-Hardness}

The reduction is from Set Cover: given a collection
$\mathcal S=\{S_1,S_2,\dots,S_s\}$ of sets where each $S_i\subseteq U$,
and given a positive integer~$k$, are there $k$ of the sets
$S_{i_1},S_{i_2},\dots,S_{i_k}$ whose union hits all elements of~$U$?

\begin{figure}
  \centering
  \includegraphics[scale=0.6]{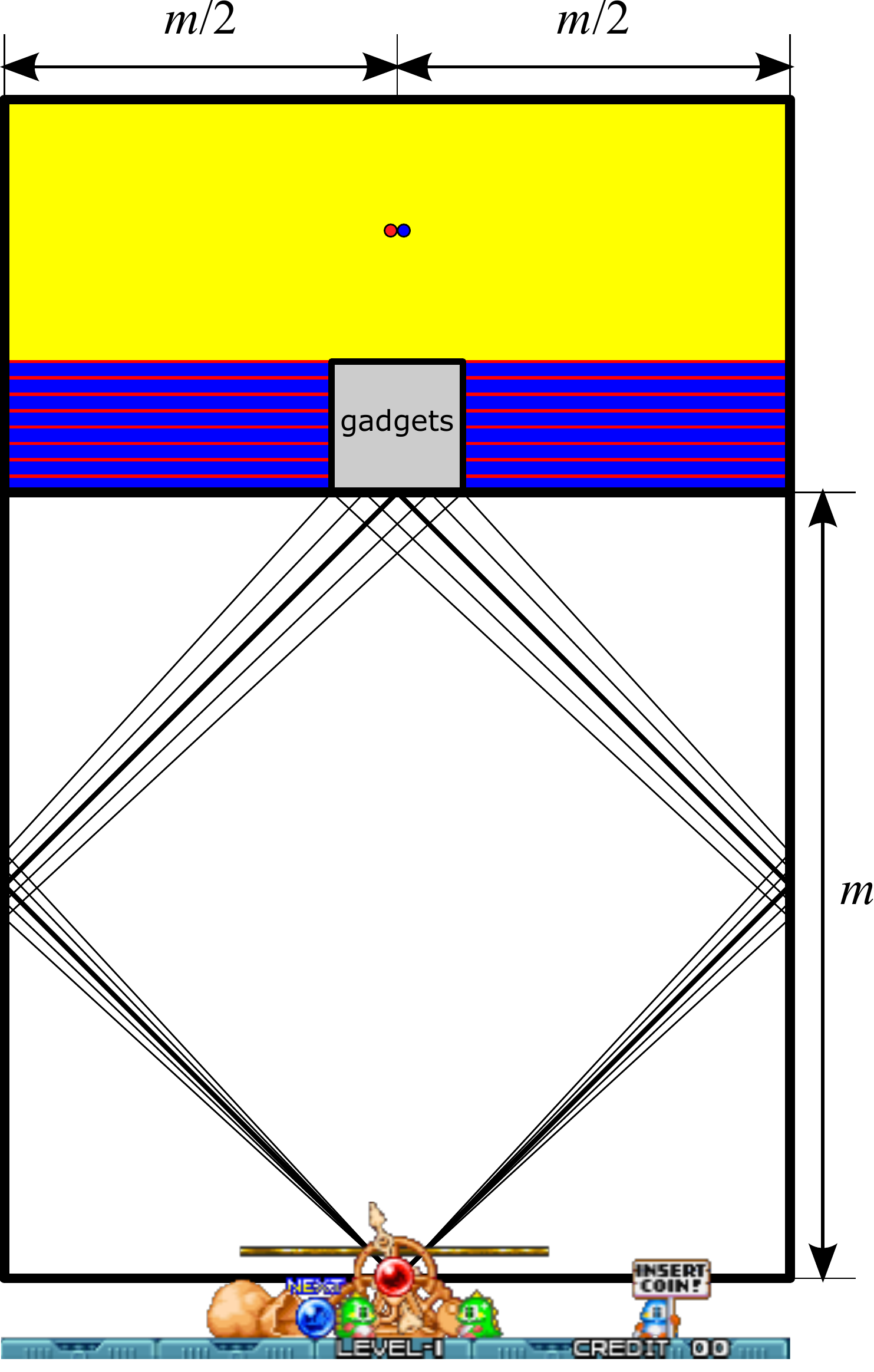}
  \caption{Overall structure of the reduction.  All other gadgets lie within a
    small square at the top of an $m \times m$ square, where $m$ is the width
    of the game.  Red horizontal lines separate the gadgets into layers,
    with blue fill in between.
    At the top is a huge rectangle of yellow bubbles with one red bubble
    and one blue bubble in the middle.}
  \label{overall}
\end{figure}

Figure~\ref{overall} shows the overall structure of the reduction.
The bulk of the construction is in the central small square, which is aligned
on the top side of an $m \times m$ square above the floor.
By making the central square small enough, the angles of direct shots at the
square are close to vertical (which we will need to solve most gadgets),
and the rebound angles that hit the square are all approximately~$45^\circ$
(which we will need to solve the crossover gadget below), even after the
ceiling falling caused by the shots in the reduction.
The player could do multiple rebounds (or destroy bubbles to cause the ceiling
to lower prematurely) to make shot angles more horizontal, but this will only
make it harder to solve the gadgets.

\subsection{Bubble Sequence}

The sequence of bubbles given to the player is as follows.
The very first color appears only $k$ times, where $k$ is the desired
set-cover size.  Each remaining color appears sufficiently many times
($\Theta(s |U|)$ times, which we will refer to as~$\infty$).
Unneeded bubbles can be discarded by forming isolated groups of
size $3$, $4$, or $5$ off to the side.

\begin{center}
  \begin{tabular}{llll}
    \phantom{$\infty$}\llap{$k$} blue, & $\infty$ yellow, & $\infty$ blue, & $\infty$ red; \\
    $\infty$ blue, & $\infty$ yellow, & $\infty$ blue, & $\infty$ red; \\
    $\infty$ blue, & $\infty$ yellow, & $\infty$ blue, & $\infty$ red; \\
    \multicolumn{1}{c}{$\vdots$} & \multicolumn{1}{c}{$\vdots$} & \multicolumn{1}{c}{$\vdots$} & \multicolumn{1}{c}{$\vdots$} \\
    $\infty$ blue, & $\infty$ yellow, & $\infty$ blue, & $\infty$ red; \\
    $\infty$ red, & $\infty$ red, & $\infty$ red, & \dots \\
  \end{tabular}
\end{center}

The rough idea is the following.
Red bubbles separate vertical layers that unravel sequentially,
as enforced by blue buffers.
Blue and yellow bubbles form triggers to communicate signals into the next layers,
alternately.
Blue triggers cup yellow triggers in the next level, and vice versa.

\subsection{Gadgets}

\begin{figure}
  \centering
  \includegraphics[scale=0.25]{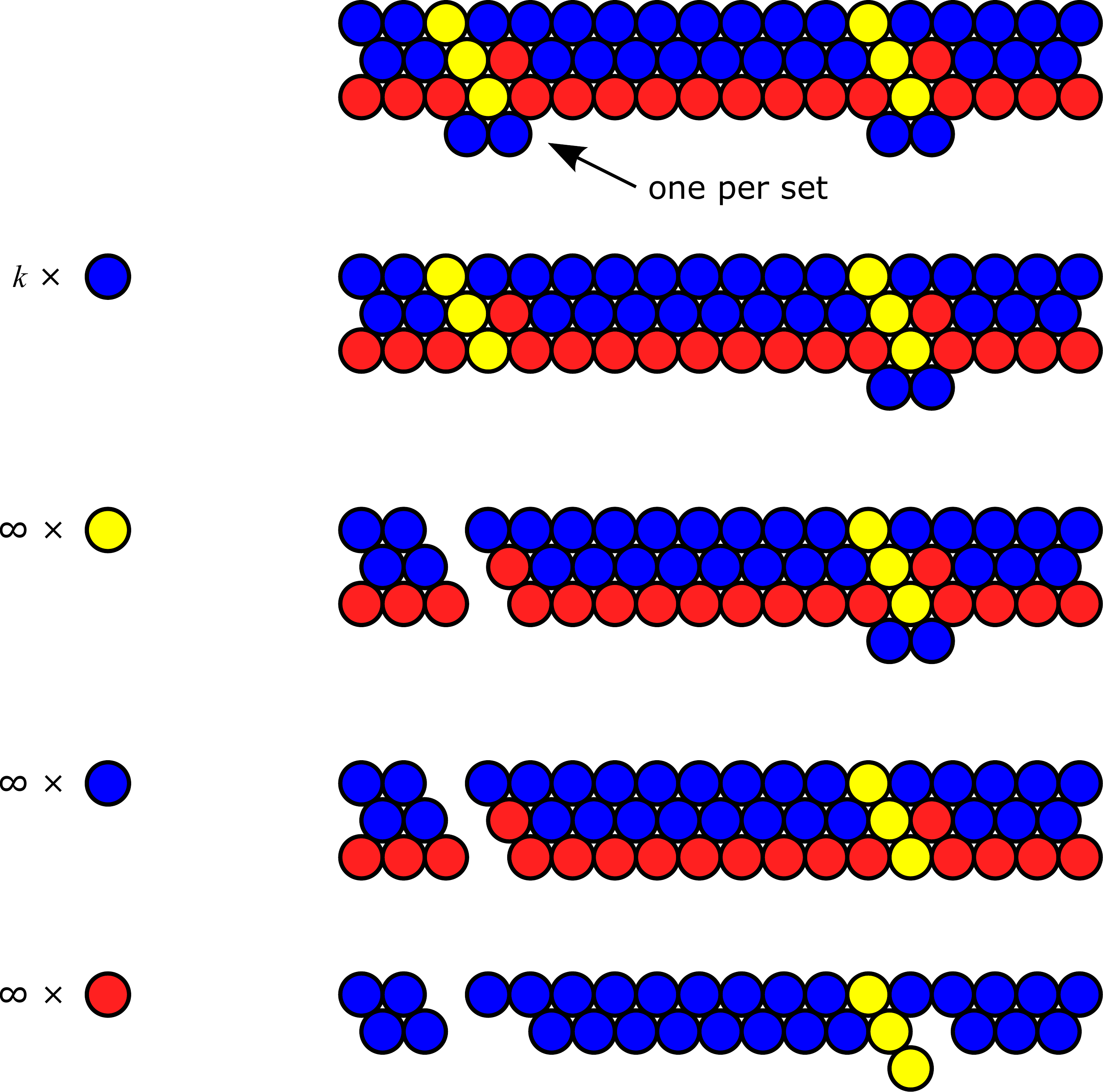}
  \caption{Choice gadget, shown here with $s=2$ sets.
    (Left) Behavior of a chosen set.
    (Right) Behavior of an unchosen set.}
  \label{choice}
\end{figure}

First we have one instance of the choice gadget, shown in Figure~\ref{choice},
which allows triggering $k$ sets (whichever the player chooses).

\begin{figure}
  \centering
  \includegraphics[scale=0.25]{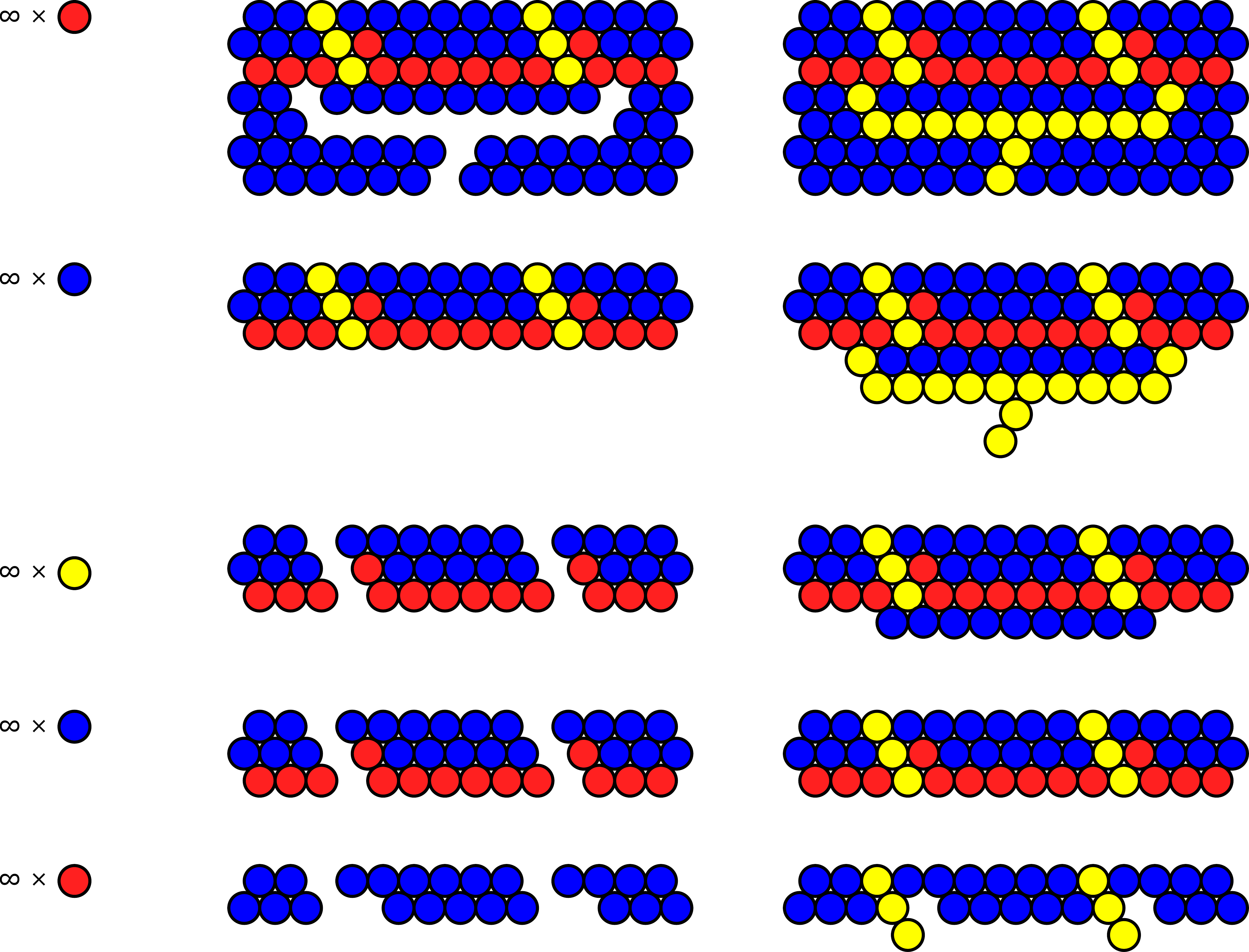}
  \caption{Split gadget.  (Left) Behavior of a chosen set.
           (Right) Behavior of an unchosen set.}
  \label{split}
\end{figure}

Then we use several split gadgets, shown in Figure~\ref{split},
to split each trigger for set $S_i$ into $|S_i|$ triggers.

\begin{figure}
  \centering
  \includegraphics[scale=0.25]{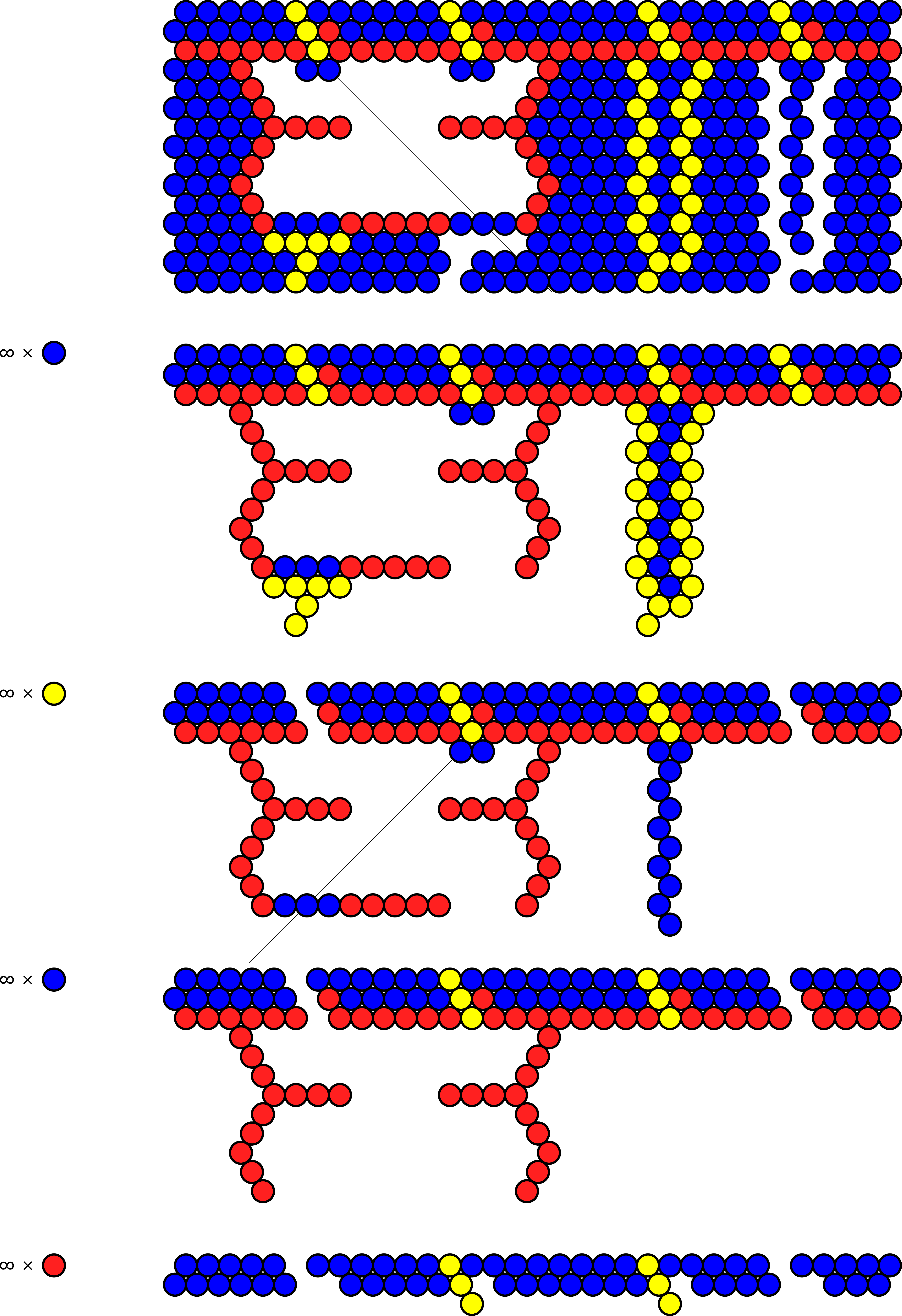}
  \caption{Crossover gadget,
     shown here with left side inactive and right side active.
     On the right are other wires whose values are simply copied.}
  \label{crossover}
\end{figure}

Then we use several crossover gadgets, shown in Figure~\ref{crossover},
to bring together all the triggers for element~$x$, for every element~$x$.
More precisely, Figure~\ref{crossover} shows how to copy all other wire values
while swapping an adjacent pair.
Adjacent swaps suffice to bubble sort $S_1,S_2,\dots,S_s$
from being in order by set to being in order by element.
In fact, we can use the parallel sorting algorithm \emph{odd--even sort}
by executing several swaps in one layer, and use only $2 \sum_i |S_i|$ layers.

\begin{figure}
  \centering
  \includegraphics[scale=0.25]{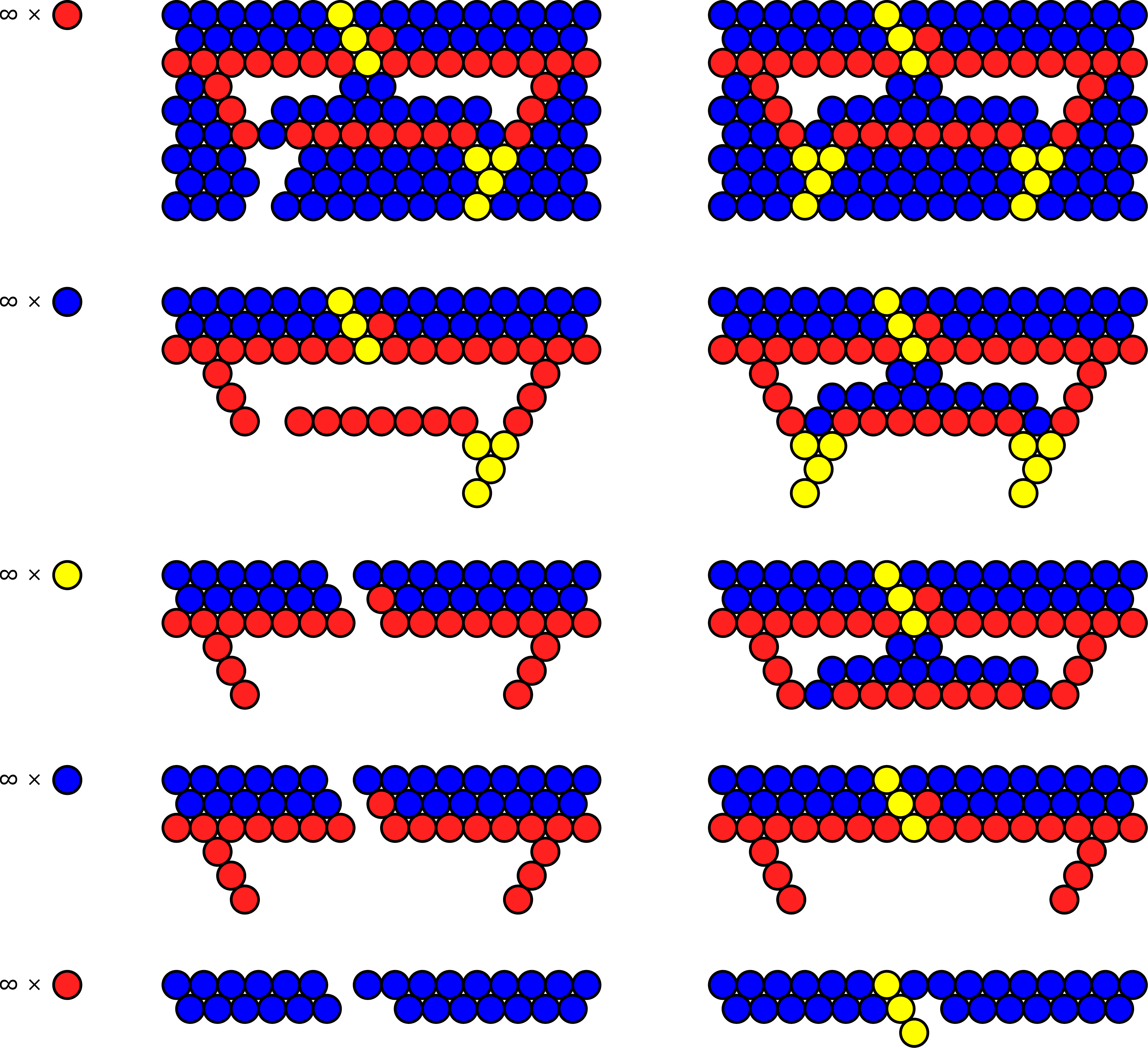}
  \caption{\textsc{or} gadget.  (Left) One input active, triggering output.
    (Right) No inputs active.}
  \label{or}
\end{figure}

Next, for each element, we merge all the triggers for that element
(coming from sets that contain the element), using the \textsc{or} gadget
in Figure~\ref{or}.
In this gadget, any input trigger enables the output trigger.
By combining several \textsc{or} gadgets, we end up with one trigger
per element in~$U$, indicating whether that element was covered by
the $k$ chosen sets.

\begin{figure}
  \centering
  \includegraphics[scale=0.25]{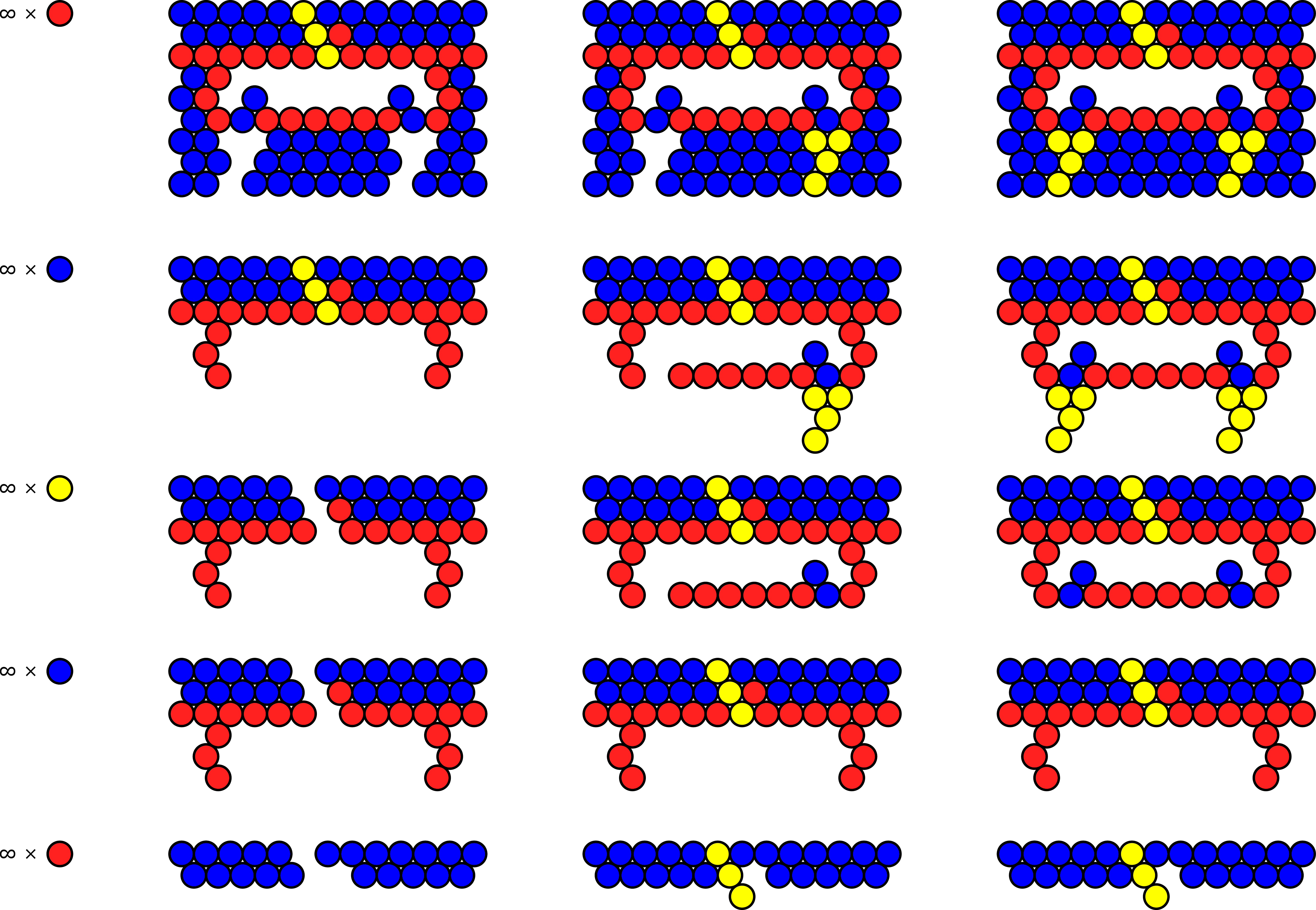}
  \caption{\textsc{and} gadget.  (Left) Two inputs active, triggering output.
    (Middle) One input active.  (Right) No inputs active.}
  \label{and}
\end{figure}

Finally, we combine the element triggers using the \textsc{and} gadget
in Figure~\ref{and}.
In this gadget, the output triggers only if all inputs trigger.
By combining several \textsc{and} gadgets, we end up with one trigger
indicating that all elements are covered, i.e., we found a set cover of
size~$k$.

This trigger is connected to a huge ($n^{1-\varepsilon}$-area) rectangle of
yellow bubbles at the top of the board, with one
red bubble in the middle, as shown in Figure~\ref{overall}.
If the yellow triggers, the player wins the game immediately (as the red falls).
Otherwise, only red bubbles come, so the player eventually dies when
the yellow rectangle reaches the floor.  (We include the red and blue bubbles
in the middle of the yellow bubbles because, in the actual game, only present
bubbles can be presented for shooting, so if there were only yellow bubbles
left, the player would get to shoot yellow and win.)

Thus, even approximating the maximum number of poppable bubbles better than a
factor of $n^{1-\varepsilon}$ is NP-hard
(similar to Tetris \cite{Breukelaar-Demaine-Hohenberger-Hoogeboom-Kosters-Liben-Nowell-2004}).

\subsection{Putting It Together}

\begin{figure}
  \centering
  \includegraphics[scale=0.25]{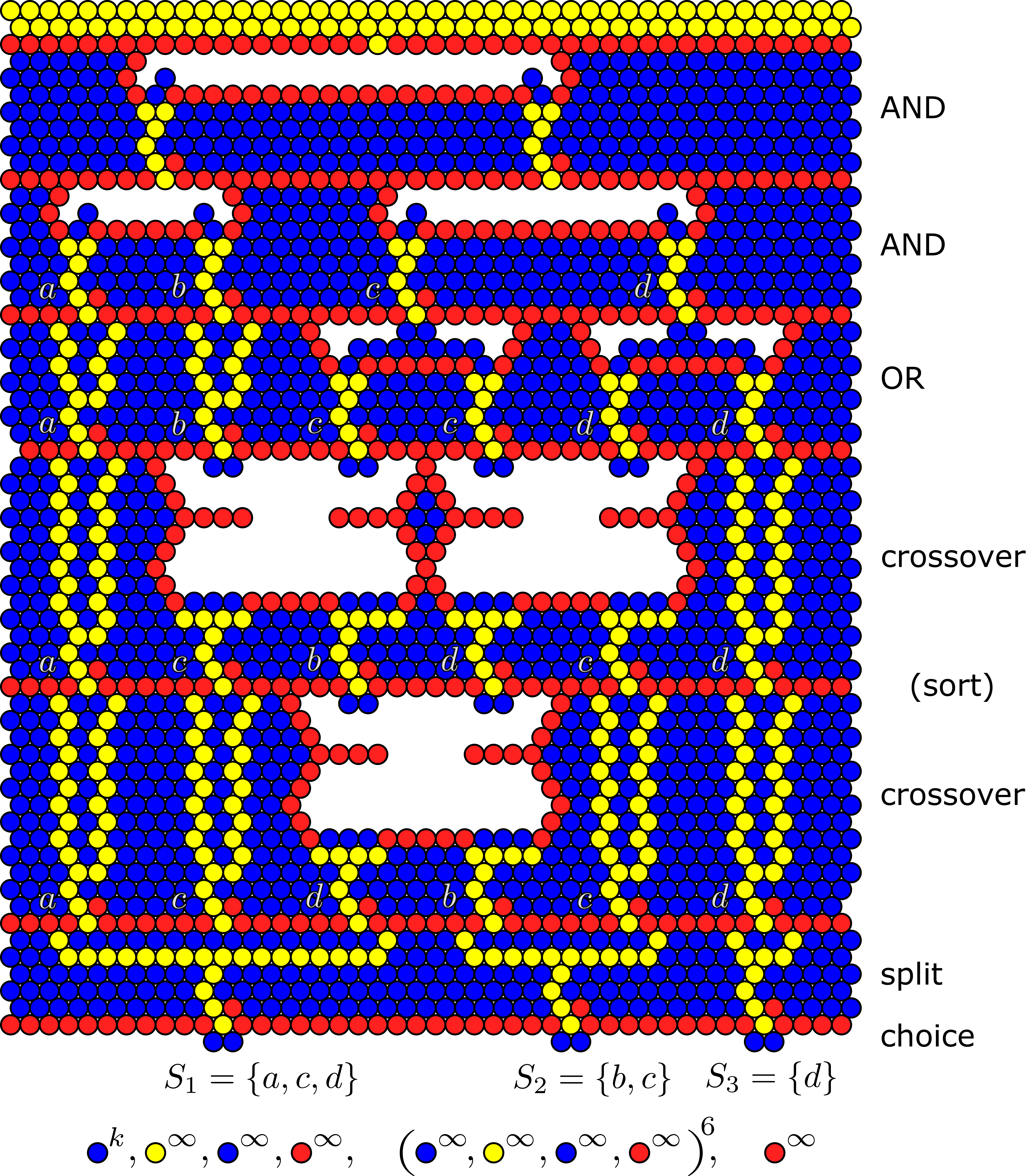}
  \caption{Example of the main construction (the gray box in
    Figure~\ref{overall}) with three sets and four elements.
    The bubble sequence at the bottom can solve the puzzle for $k=2$ and $k=3$,
    but not for $k=1$.}
  \label{example}
\end{figure}

\begin{figure}
  \centering
  \includegraphics[scale=0.25]{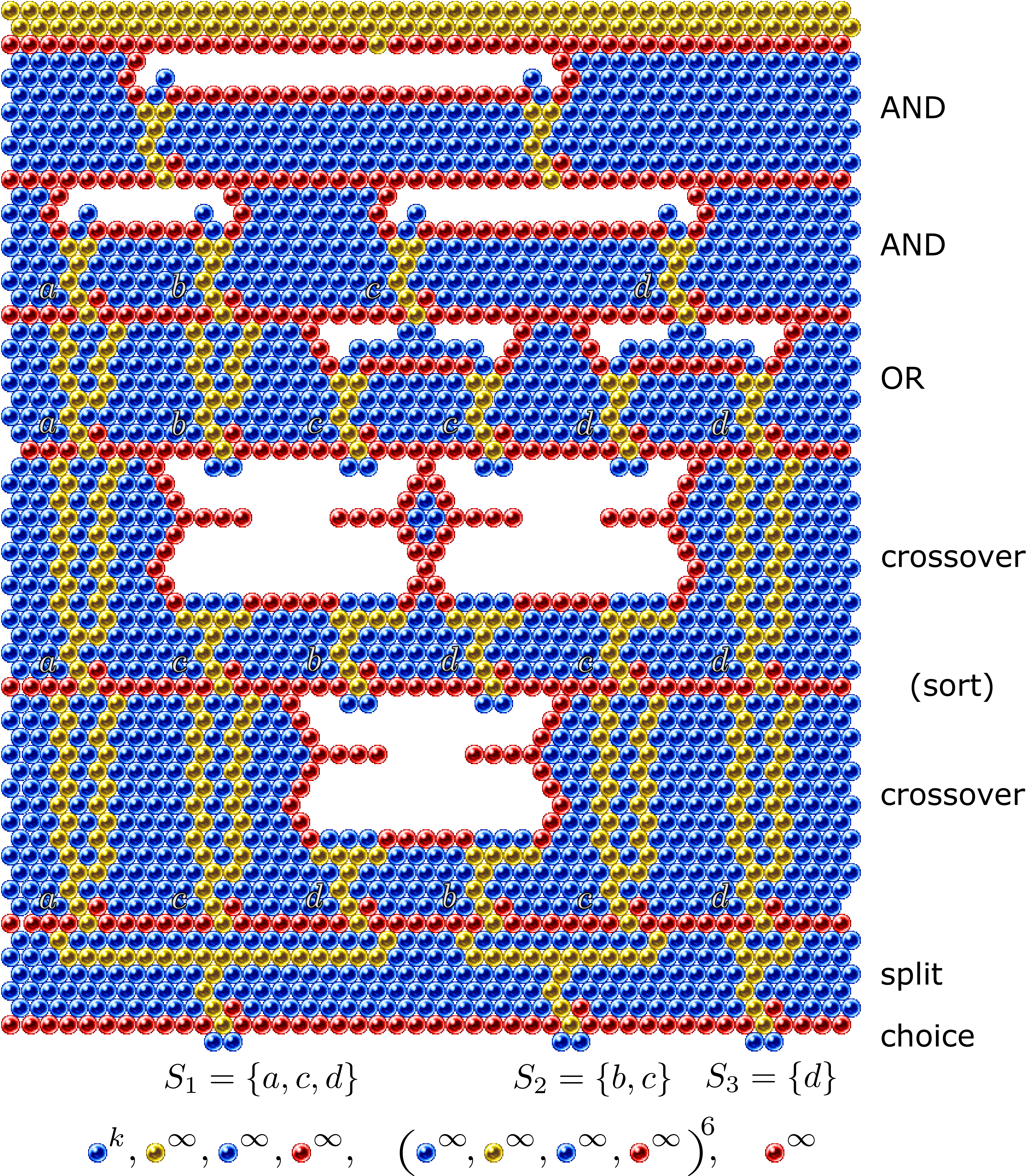}
  \caption{Figure~\ref{example} using actual Puzzle Bobble sprites,
    thanks to The Spriters Resource.}
  \label{example pixel}
\end{figure}

Figures~\ref{example} and~\ref{example pixel}
show an example of how the gadgets fit together in a real example.
In particular, it illustrates how to stretch gadgets horizontally so that
their inputs and outputs align, and how to stack the layers of gadgets
(each gadget is placed on the row immediately after the previous).

The bijection between solutions of the Puzzle Bobble instance and the
Set Cover instance come from which triggers get popped by the first $k$ blue
shots on the Choice gadget.  (Fewer than $k$ triggers could be popped,
corresponding to smaller-than-$k$ set covers.)  The correctness follows from
the claimed properties of the gadgets, which can be verified from the figures
implementing a greedy algorithm of popping all possible bubbles of each
provided color (which can only help for these instances).

A key lemma for correctness is that, during the $i$th blue--yellow--blue--red
phase of the bubble sequence, only bubbles in the $i$th layer of the
construction can be directly popped, with spillover into the next layer only
from triggered yellow bubble wires.  The $i$th red layer prevents any nonred
bubbles from physically reaching the next layer in the $i$th phase, because
the gaps between red bubbles are designed to be strictly less than one bubble
width.  (Precisely, the gap width is $\sqrt 3 - 1 \approx 0.73$.)
At the end of the phase when firing red bubbles, the blue in the next layer
uses the same $<1$ gaps (when the yellow has been triggered) to prevent any red
bubbles from reaching the next layer.  So the lemma follows.

\section{Open Problems}

We have proved NP-hardness for just three colors.
What about just two colors?  Or even one color?

\section*{Acknowledgments}

We thank Giovanni Viglietta for helpful discussions, in particular for
pointing out bugs in earlier versions of this proof.

\let\realbibitem=\bibitem
\def\bibitem{\par \vspace{-1.2ex}\realbibitem}

\bibliography{combinatorialgames}
\bibliographystyle{alpha}

\end{document}